\documentclass[a4paper,20pt]{article}
    \usepackage[T1]{fontenc}
    \usepackage{graphicx}
    \usepackage{amsmath}
    \usepackage{amsfonts}
    \renewcommand{\abstract}{}
\begin{document}
\makeatletter
\renewcommand{\@oddhead}{\textit{YSC'14 Proceedings of Contributed Papers} \hfil \textit{A. Martin-Carrillo, M. Kirsch,
E. Kendziorra, R. Staubert}}
\renewcommand{\@evenfoot}{\hfil \thepage \hfil}
\renewcommand{\@oddfoot}{\hfil \thepage \hfil}
\fontsize{11}{11} \selectfont

\title{X-ray Timing Analysis of Six Pulsars Using ESA's \emph{XMM-Newton} Observatory}
\author{\textsl{A. Martin-Carrillo$^{1}$, M. Kirsch$^{1}$,
E. Kendziorra$^{2}$, R. Staubert$^{2}$}}
\date{}
\maketitle
\begin{center} {\small $^{1}$European Space Astronomy Centre (ESAC) ESA,
Apartado P.O. Box 50727, 28080 Madrid, Spain  \\
$^{2}$Institut f\"ur Astronomie und Astrophysik der Universit\"at
T\"ubingen, Sand 1, 72076 T\"ubingen, Germany \\
acarrillo@sciops.esa.int}
\end{center}

\begin{abstract}
We present results of a timing analysis of various isolated pulsars
using ESA's \emph{XMM-Newton} observatory. Isolated pulsars are
useful for calibration purposes because of their stable emission. We
have analyzed six pulsars with different pulse profiles in a range
of periods between 15 and 200 ms. All observations were made using
the \emph{EPIC-pn camera} in its faster modes (Small window, Timing
and Burst modes). We investigate the relative timing accuracy of the
camera by comparing the pulse periods determined from the
\emph{EPIC-pn camera} observations with those from radio
observations. As a result of our analysis we conclude that the
relative timing accuracy of the \emph{EPIC-pn camera} is of the
order of $1\times 10^{-8}$.

\end{abstract}

\section*{Introduction}
\indent \indent The \emph{EPIC-pn camera} aboard \emph{XMM-Newton}
is designed to read out data in different modes with respect to time
and spatial resolution. For the timing analysis we use the so called
Small Window, Timing and Burst modes, with time resolutions of 6 ms,
0.03 ms and 7 $\mu$s, respectively. Using these modes and a number
of isolated pulsars as calibration targets we investigate the
relative timing capabilities of the \emph{EPIC-pn camera}.
\par
Isolated pulsars are characterized by stable pulse periods and
pulse profiles making them useful calibration targets
(even under the existence of irregularities like glitches and timing noise).
For our timing analysis we use photons from 0.2 keV to 15 keV.
\par
Here we present an analysis of the relative timing accuracy of the
EPIC-pn camera onboard \emph{XMM-Newton} using six different pulsars
with periods between 15 and 200 ms. We have selected pulsars with a
variety of different pulse profiles in order to see whether the
timing results depend on pulse shape.

\section*{Observations and data analysis}
\indent \indent We have analyzed different pulsars observed by
\emph{XMM-Newton} between 2000 and 2007. As the main calibration
source for timing, the Crab pulsar is regularly observed twice per
year (spring and autumn) to make sure that the measurements do not
depend on the Earth's orbital phase. The exposure times of the
observations analyzed are between 3 and 50 ks. The observations were
mainly done in Timing or Burst mode, except those of the Vela pulsar
which were done in Small Window mode. In Figure~1 we show the pulse
profiles of all pulsars studied.

We define the relative timing accuracy with reference to the highly
accurate measurements by radio telescopes trough expression (1),
where $P_{X}$ is the X-ray period calculated by us and $P_{R}$ is
the radio period extrapolated or interpolated from data found in
radio pulsar databases (e.g. the Jodrell Bank Observatory
\footnote{http://www.jb.man.ac.uk/~pulsar/crab.html}, ATNF, European
Pulsar Network, or others).
\begin{equation}
\centering \dfrac{\Delta{P}}{P}=\dfrac{P_{X}-P_{R}}{P_{R}}
\end{equation}
\par
For the Crab pulsar, the Jodrell Bank pulsar group provides monthly
ephemeris of the pulsar to the scientific community.
Through linear interpolation we can estimate the radio
period for the time at which the X-ray observation were made.
Assuming a negligible uncertainty in the radio period
the $\Delta$P obtained is the error of our analysis.
For the other pulsars we have less complete information, such that in general
only an extrapolation to the time of the X-ray observation is possible,
leading to a less accurate period estimate.
Glitches between the radio observation and the X-ray observation can
dramatically change the ephemeris and ruin our analysis.
Therefore we have taken great care to find the closest ephemeris.
For pulsars like the Crab pulsar or Vela
pulsar an ephemeris less than one months ago needs to be taken, whereas for
other more stable pulsars an ephemeris from even years ago may be used.

The extrapolated or interpolated radio period is then used for a first
trial in the search of the X-ray period. This search is made using
the \textit{Xronos} routine called \textit{efsearch}.
\textit{efsearch} is performing an epoch-folding with a range
of trial periods, calculating the $\chi^{2}$ sum which describes the
deviation of the resulting profile from a flat distribution. The period
at which $\chi^{2}$ is maximum is considered to be the correct pulse period.
In practice we do not use the maximum, but rather the weighted mean of
the $\chi^{2}$ distribution from a fit with a triangular or a Gaussian
profile.
In Figure~2 we show
$\chi^{2}$ distributions of the period search for one observation
of each pulsar. The width of the distribution
is different for each pulsar, depending on the pulse period
and the elapsed time of observation (as discussed below).
\par
As mentioned earlier, we consider that the value of the radio
period calculated on the epoch of the X-ray observation
has no error, so the difference between the radio period and
our X-ray period $\Delta{P}$ is taken as the error of
our measurement. We can approximate
the $\chi^{2}$ distribution by a triangle where the maximum
corresponds to the true period P$_{0}$ and the points P$_{1}$
and P$_{2}$ where the legs of the triangle meet the level of constant
$\chi^{2}$ define the total width of the $\chi^{2}$ distribution.
For a pulse profile with a small single peak, P$_{1}$ and P$_{2}$
can be calculated by expression (2), where T$_{obs}$ is the elapsed
observational time and N$_{per}$ is the number of pulse periods in this
time.
\begin{equation}
\centering P_{1}=\dfrac{T_{obs}}{N_{per}+1}; \;\; P_{2}=\dfrac{T_{obs}}{N_{per}-1}
\;\;\; where \; N_{per}=\dfrac{T_{obs}}{P}
\end{equation}

For a triangular function the Full Width Half Maximum (FWHM)
is equal to $(P_{1} - P_{2})/2$ and can be expressed as in
expression (3) as function of the period and the elapsed
observation time.
\begin{equation}
\centering FWHM=\dfrac{P_{2}-P_{1}}{2} \Rightarrow
FWHM=\dfrac{P^{2}}{T_{obs}}
\end{equation}

We use this expression to predict the FWHM of the
$\chi^{2}$ distribution, which also serves as an upper limit of the
uncertainty of the measured period. Empirically we have found that
a rough estimate of the uncertainty of the measured period can be
found by dividing the FWHM by the number of phase bins used to
construct the pulse profile.
We have generally used 10 bins for the pulse profiles, except
in the case of the Crab pulsar where we used 100 bins.
In Fig.~2 (right) one line is shown for each pulsar giving the predicted
FWHM as calculated by expression (3) normalized to the period P.
All the values of FWHM/P measured are close to the
predicted ones.

\section*{Results}
\indent \indent We have measured the relative timing accuracy of the
\emph{EPIC-pn camera} for six pulsars (see Fig.~1), most frequently
for the Crab pulsar. As we have mentioned earlier \emph{XMM-Newton}
performs two observation of the Crab every year, such that we were
able to analyze 25 observations (of duration between $2\times
10^{3}$ and $4\times 10^{4}$ sec each). For the Vela pulsar we have
four observations, for PSR B1509-58 two observations, and for the
remaining two pulsars only one observation each.
\par
In Fig.~3 (top/left) we compare the estimated timing accuracy
with the actually observed one. There is one line for each pulsar representing
the estimated relative accuracy as a function of observing time (using the
FWHM as calculated by expression (3) and divided by the number of bins used
to construct the pulse profiles). The observed accuracy, represented by the
data symbols, are from the difference between the periods determined in the
X-rays from observations by \emph{XMM-Newton} and the radio periods (taking the
absolute value).
We find that all observed data points are below the lines of the
estimated accuracies, except two: one corresponding to the Vela pulsar,
the other to PSR B1509-58. We find that in both cases, the used radio
periods appear unreliable since they were determined far away from
the time of the X-ray observations (more than one year for the Vela pulsar
and about five years for PSR B1509-58). So, we exclude those two points from the
following discussion. Fig.~3 (top/right) shows the absolute $\Delta$P/P as a
function of observing date. There is no obvious change of the timing accuracy
of the \emph{EPIC pn-camera} over its lifetime.
\par
In the lower two panels of Fig.~3 only results from observations
of the Crab pulsar are shown. Again, there is no obvious dependence on date,
but there is a tendency of a smaller uncertainty for longer observations,
as would be expected.
\par
For a quantitative measure of the timing accuracy we use the standard
deviation of the distributions of $\Delta$P/P values (shown in Fig.~3).
Fitting the distributions with a Gaussian normal distribution we
find a standard deviation of $7\times 10^{-9}$ for all pulsars (including
the Crab pulsar) and $5\times 10^{-9}$ for the Crab pulsar alone. While
the distribution for Crab pulsar values is centered at zero (within
uncertainty) the mean value of the distribution for all data is slightly
offset, in the sense that the X-ray period is slightly larger on
average than the radio period (this is due to data from pulsars other
than the Crab pulsar).

\section*{Conclusions}
\indent \indent We have determined X-ray pulse periods by epoch
folding of six different pulsars (with periods between 15 and 200
ms), including the Crab pulsar from observations by the
\emph{EPIC-pn camera} of \emph{XMM-Newton}. By comparing the X-ray
periods with inter-(extra-)polated radio periods from public
archives we find generally very good agreement (except in three
cases where the radio periods were taken far away from the time of
the X-ray observations). Under the assumption that the radio periods
have no uncertainty, the difference between the X-ray and radio
periods give an estimate of the accuracy of the X-ray measurements.
We conclude that (for integration times of a few ks) the relative
timing accuracy of the \emph{EPIC-pn camera} is generally better
than $1\times 10^{-8}$.
\par
Further analysis on the timing accuracy of \emph{XMM-Newton's} \emph{EPIC-pn
camera} will be presented in M. Kirsch et al. 2007 (in preparation).

\section*{Acknowledgments}
\indent \indent The \emph{XMM-Newton} project is an ESA science
mission with instruments and contributions directly funded by ESA
Member States and the USA (NASA). A. Martin-Carrillo would like to
thank the ESAC Faculty group for their financial support.

\newpage

\textbf{Figure 1.} Pulse profiles of the different pulsars analyzed.
From top left to bottom right these are: PSR J0537-69 (16 ms), Crab
pulsar (33 ms), PSR B0540-69 (50 ms), Vela pulsar (89 ms), PSR
B1509-58 (151 ms) and PSR B1055-52 (197 ms). \vspace{10ex}

\textbf{Figure 2.} Left: $\chi^{2}$ distributions for one
observation each of the studied pulsars. Right: comparison of the
predicted relative width of the $\chi^{2}$ distributions FWHM/P
(lines) and the observed ones (symbols).\vspace{10ex}

\textbf{Figure 3.} The relative timing accuracy of
\emph{XMM-Newton's} \emph{EPIC-pn camera}. Top/left: comparison of
the expected absolute($\Delta$P/P) using the predicted FWHM divided
by the number of phase bins used to construct the pulse profiles
(lines) and the observed relative uncertainty by comparing our X-ray
period with the extra-(inter-)polated radio period (symbols). The
number of bins in the pulse profiles used is generally 10, except
for the Crab pulsar, for which we used 100 bins. Top/right:
absolute($\Delta$P/P) for all pulsars as a function of date in MJD.
The lower panel gives absolute($\Delta$P/P) for Crab pulsar data
only: as a function of date in MJD (left), and as a function of
observational time in ks (right).\vspace{10ex}

Figures are available on YSC home page
(http://ysc.kiev.ua/abs/proc14$\_$5.pdf).

\end{document}